\documentstyle[12pt,aaspp4,psfig]{article}

\lefthead{Spaans \& Neufeld}
\righthead{H$_2$ in 3-D Diffuse Clouds}

\begin{document}

\title{Molecular Hydrogen in Diffuse Interstellar Clouds \\
       of Arbitrary Three-Dimensional Geometry}

\author{Marco Spaans and David A. Neufeld}
\affil{Department of Physics \& Astronomy,
Johns Hopkins University, 3400 North Charles Street,
Baltimore, MD 21218-2686}

\begin{abstract}

We have constructed three-dimensional models for the equilibrium
abundance of molecular hydrogen within diffuse interstellar
clouds of arbitrary geometry that are illuminated by ultraviolet
radiation.  The position-dependent photodissociation rate of H$_2$
within such clouds was computed using a 26-ray approximation to
model the attenuation of the incident ultraviolet radiation field
by dust and by H$_2$ line absorption.

We have applied our modeling technique to the isolated diffuse
cloud G236+39, assuming that the cloud has a constant density
and that the thickness of the cloud along the line of sight is
at every point proportional to the 100 $\mu$m continuum intensity
measured by IRAS.  We find that our model can successfully
account for observed variations in the ratio of 100 $\mu$m continuum
intensity to HI column density, 
with 
larger values of that ratio occurring
along lines of sight in which the molecular hydrogen fraction is
expected to be largest.

Using a standard $\chi^2$ analysis to assess the goodness
of fit of our models, we find (at the 60$\sigma$ level)
that a three-dimensional model is more successful in matching
the observational data than a one-dimensional model
in which the geometrical extent of the cloud along the line of sight
is assumed to be 
very much smaller than its extent in the plane-of-the-sky.
If $D$ is the distance to G236+39,
and given standard assumptions about the rate of grain-catalysed 
H$_2$ formation,
we find that the cloud has an extent along the line of sight
that is $0.9 \pm 0.1$ times its mean extent projected onto
the plane of the sky; a gas density of $(53 \pm 8) \,(100\,{\rm pc}/D)$
H nuclei per cm$^{3}$; and is illuminated by a radiation field of
$(1.1 \pm 0.2)\,(100\,{\rm pc}/D)$ times the mean interstellar radiation field
estimated by Draine (1978).  The derived 100 $\mu$m emissivity per
nucleon is $(1.13 \pm 0.06)\times 10^{-20} \rm \,MJy\,sr^{-1} cm^2$.

\end{abstract}


\section{Introduction}

Diffuse interstellar clouds of visual extinction
$A_{\rm V}\sim 1$ mag represent the intermediate
case between {\em atomic clouds} in which the molecular
fraction is very small and {\it dense molecular clouds} in which
gas is almost fully molecular; as such, they represent
the best laboratories in which to study the formation and
destruction of molecular hydrogen.

\subsection{Observations of molecular hydrogen in diffuse clouds}

Direct observations of molecular hydrogen in diffuse molecular
clouds have been severely hampered by the absence of any
dipole-allowed transitions at radio, infrared or visible wavelengths.
Given current instrumental sensitivities, 
direct measurements of the H$_2$ abundance in diffuse clouds
can only be obtained from observations of the {\it dipole-allowed}
electronic transitions of the Lyman and Werner bands.  These
transitions, which lie in the $912-1100$ \AA\ spectral region,
were detected by the {\it Copernicus} satellite in absorption
toward more than one hundred background sources (Spitzer et al.\ 1973;
Savage et al.\ 1979).  Although the planned
{\it Far Ultraviolet Spectroscopic Explorer} (FUSE) will allow
such absorption line studies to be carried out toward fainter
background sources than those observed with {\it Copernicus},
direct observations of molecular hydrogen will continue to
allow only sparse spatial sampling of diffuse molecular clouds.

An alternative -- although less direct -- means of studying the
distribution of H$_2$ in diffuse molecular clouds is afforded
by the combination of HI 21cm observations and 100 $\mu$m
continuum observations carried out by the IRAS satellite.  The
diffuse cloud that has been studied most extensively by this
method is G236+39, an isolated high-latitude cloud 
(Galactic coordinates $b=38.23$, $l=235.73$) of angular size $\sim 2 \times 1$ 
degrees for which Reach, Koo \& Heiles (1994) have used the Arecibo radio
telescope to obtain a complete HI map at a spatial resolution (3 arcminutes) 
comparable to that of the IRAS beam ($3 \times 5$ arcminutes).
Given the absence of any stellar sources associated with G236+39,
the distance is unfortunately not well determined.   
Figure 1, based upon data kindly provided in
electronic form by W.~Reach, shows the HI column densities,
$N({\rm HI})$, and 100 $\mu$m continuum intensities, $I_{100}$, for more
than 4000 lines of sight toward this cloud.  Assuming that $I_{100}$
is linearly proportional to the column density of H nuclei, $N_{\rm H}=
N({\rm HI}) + 2\,N({\rm H_2})$, Reach et al.\ argued that
the higher ratios of $I_{100}/N({\rm HI})$ typically observed when
$N({\rm HI})$ is large represented the effects of molecule formation
along the line of sight.  This allowed the H$_2$ column density
to be estimated as $N({\rm H_2})=0.5\,[(I_{100}/r_0) - N({\rm HI})]$,
where $r_0 = I_{100}/N_{\rm H}$ is the 100 $\mu$m emission
coefficient per H {\it nucleus},
a quantity that could be determined from the slope of the $I_{100}$ versus
$N({\rm HI})$ correlation in the small column density limit where the molecular
fraction was negligible.

\subsection{Models of diffuse clouds}

The chemistry of diffuse molecular clouds has been the subject
of extensive theoretical study over the past 25 years
(e.g. Black \& Dalgarno 1977; van Dishoeck \& Black 1986;
Spaans 1996, and references therein).  Common to all theoretical
models for diffuse molecular clouds are the assumptions that
the formation of molecular hydrogen takes place by means of
grain surface reactions and that the destruction of H$_2$
occurs primarily by photodissociation.  Because photodissociation
of H$_2$ follows {\em line absorption} of ultraviolet radiation in the Lyman
and Werner bands, the photodissociation rate is diminished
by self-shielding and shows a strong depth-dependence.

Most previous studies of diffuse cloud chemistry have assumed
a one-dimensional plane-parallel geometry, in which a geometrically-thin
slab is illuminated by ultraviolet radiation from either one or two
sides.  The solid curve in Figure 1 shows the predicted variation of
$I_{100}$ with $N({\rm HI})$ for a series of 
slab models illuminated from two sides 
with a constant assumed density 
of H nuclei 
and a variable column density.
As described in \S 3 below, the assumed 100$\mu$m emissivity per
H nucleus and the cloud density have been chosen to optimize
the fit to the data.   In the limit of small $N({\rm HI})$,
the predicted relationship between $I_{100}$ and $N({\rm HI})$ is linear,
but for $N({\rm HI}) \ge \rm few \, \times 10^{20} cm^{-3}$,
the molecular hydrogen fraction becomes significant and the curve
turns upwards.

While a one-dimensional slab model can evidently account for the
gross properties of the data plotted in Figure 1, we have investigated
in this study whether a {\em better} fit to the data can be obtained
with models in which a {\em three-dimensional} cloud structure is assumed.
Thus instead of modeling the cloud G236+39 as a ``cardboard cutout'' that
has 
an extent along the line of sight that is very much smaller than
its extent in the plane-of-the-sky, we have constructed
a series of three-dimensional models in which we assume the
cloud to have a significant geometrical thickness perpendicular 
to the plane of the sky.
In other words, we include the photodissociative effects of radiation
that is incident upon the cloud from the sides as well as along
the line of sight. 

This paper is organized as follows. In \S 2 we describe the numerical method
used to model the formation and dissociation of H$_2$ in a three-dimensional
diffuse cloud of arbitrary shape.  In \S 3 we investigate
whether our three-dimensional models describe more accurately than
one-dimensional models the
correlation between IRAS 100 $\mu$m emission
and HI column density for the line of sight toward the cirrus cloud
G236+39.  The results of our investigation are discussed in \S 4.

\section{Computational Method}

We have constructed a model for the equilibrium H$_2$ abundance within a
three-dimensional molecular cloud 
with a constant density of H nuclei
and an arbitrary shape.
This study extends the spherical cloud models described by Neufeld \& Spaans
(1996).  

We assume that molecular hydrogen is formed catalytically on dust
grains at a rate $R\,n_{\rm H} n({\rm HI})$ per unit volume, where
$n({\rm HI})$ is the density of hydrogen atoms,
$n_{\rm H}=n({\rm HI})+2\,n({\rm H_2})$ is the density of hydrogen nuclei,
and $R= 3 \times 10^{-17} \,\alpha\,\rm cm^3\,s^{-1}$ is an effective
rate coefficient with a ``canonical value" of
$3 \times 10^{-17} \rm cm^3 \, s^{-1}$ (Jura 1974) corresponding to $\alpha=1$.

The destruction of H$_2$ is dominated by spontaneous radiative
dissociation, following the absorption of 912--1110 \AA\
ultraviolet radiation in the Lyman and Werner bands.  Given the
estimate of Draine (1978) for the mean UV interstellar radiation field,
and the Lyman and Werner band oscillator strengths and
photodissociation probabilities of Allison \& Dalgarno (1970) and
Stephens \& Dalgarno (1972),
the photodissociation rate for an unshielded H$_2$ molecule is
$\zeta_0 = 5.0 \times 10^{-11} \rm \,s^{-1}$.
Once the Lyman and Werner transitions become optically-thick, the
photodissociation rate is diminished due to self-shielding.
In modeling the decrease of the photodissociation
rate due to self-shielding along any given ray, we adopt the
expression given by Tielens \& Hollenbach (1985)
for the self-shielding function for spectral lines with a Voigt
profile. 
We assume that dust absorption decreases the photodissociation rate further by a
factor $\exp(-2.55A_{\rm V})$, where $A_{\rm V}$ is the visual extinction
along the ray.
The value for the exponent, $-2.55A_{\rm V}$, is based upon the
opacity, albedo and
scattering behavior of interstellar grains given by Roberge et al.~(1991),
and has been verified with the spherical harmonics
method of Roberge (1983) for a plane-parallel slab and with the
Monte Carlo code of Spaans (1996) for a sphere.

With these assumptions, the H$_2$ photodissociation rate, $\zeta$,  at an
arbitrary point P within the cloud may be written
$${\zeta \over \zeta_0} =
{1 \over 4 \pi} \int I_{UV}({\bf {\hat k}}) \,  f(N[{\rm H}_2,{\bf {\hat k}}])
\, \exp(-2.55\,A_{\rm V} [{\bf {\hat k}}]) d\Omega, \eqno(1)$$
where $N[{\rm H}_2,{\bf {\hat k}}]$ is the column density of H$_2$
molecules along direction ${\bf {\hat k}}$ from the cloud surface
to point P, $A_{\rm V} [{\bf {\hat k}}]$ is the visual extinction
along the same path, $f(N)$ is the H$_2$ self-shielding function,
and $I_{UV}({\bf {\hat k}})$ is the incident ultraviolet intensity in
direction $({\bf {\hat k}})$ in units of the mean interstellar radiation
field given by Draine (1978).  
In the absence of any reliable distance estimate for the cloud
to which our computational method will be applied (G236+39), 
we have no means of estimating the angular dependence of $I_{UV}$.  
We therefore make the simplest assumption that 
the ultraviolet radiation field is isotropic, 
$I_{UV}({\bf {\hat k}}) = I_{UV}$.
The equilibrium H$_2$ abundance at point P is given by
$${n({\rm H}_2) \over n_{\rm H}} = {R\,n_H \over \zeta + 2\,R \,n_H}.\eqno(2)$$
The H$_2$ abundance is a decreasing function of the parameter
$I_{UV}/(\alpha n_{\rm H})$, which characterizes the relative rates of
H$_2$ photodissociation and H$_2$ formation.

In our computer model for the formation and photodissociation of H$_2$
in three-dimensional clouds, the angular integration in equation
(1) was performed using a 26-ray approximation.  Here the angular integration
was discretized on a grid of 26 directions, described in Cartesian
coordinates by the {\bf k}-vectors $(x,y,z)$ 
with $x$, $y$, and $z$ taking any of the
values --1, 0, or 1.   The H$_2$ abundance was computed at every point
using equations (1) and (2).  Since the shielding column densities
$N[{\rm H}_2,{\bf {\hat k}}]$ that enter equation (1) depend upon the
molecular abundances obtained in equation (2), an iterative method
was used to obtain a simultaneous solution, with the convergence criterion
that the change in H$_2$ abundance between successive iterations was
everywhere less than 1\%.

To verify the accuracy of our computer code, one-dimensional plane-parallel
slab calculations were performed and compared with results obtained
from the 1-D computer code of
Black \& van Dishoeck (1987).  Figure 2 shows a representative computation
for a model slab with $T=50$ K, $I_{\rm UV}=3$, $n_{\rm H}=200$ cm$^{-3}$
and $A_{\rm V}=1$ mag.  The predicted H$_2$ abundances agree to within 2\%.

\section{The IRAS Cloud G236+39}

The IRAS cloud G236+39 is an isolated, high latitude 
($b=38.23$)
cirrus cloud that provides an excellent test case in which our
three-dimensional diffuse cloud models can be assessed. 
The absence of any stellar sources associated with G236+39 causes
the distance to be not well determined.

\subsection{Assumed geometry}

As described in \S 1 above, Reach et al.\ (1994) have
used the Arecibo radiotelescope to map the HI 21cm line emission
from G236+39 at an angular
resolution of 3 arcminutes, yielding estimates of the HI column density
along 4421 lines of sight through this cloud.
We obtained estimates of the IRAS 100 $\mu$m continuum intensity
radiated by G236+39 along the same lines of sight, using the IRSKY facility
made available by IPAC\footnote{The Infrared Processing and Analysis
Center is funded by the National Aeronautics and Space Administration (NASA) as
part of the IRAS extended mission program under contract to the Jet Propulsion
Laboratory, Pasadena, USA.} to estimate and subtract the 100 $\mu$m
background.

We then interpreted the background-subtracted 100 $\mu$m intensity map
as a ``topographical contour map'' of the cloud, assuming that the intensity
at each position is proportional to the spatial extent of the cloud along
the line of sight.  Two critical assumptions underlie this interpretation:
we assume (1) that the 100 $\mu$m intensity is linearly proportional
to the column density of H nuclei (Boulanger \& P\'erault 1988), i.e.
that $I_{100} = r_0\,N_{\rm H}$ with $r_0$ constant; and (2) that the density
of H nuclei within the cloud, $n_{\rm H}$, is also constant.
With these assumptions, the linear extent along the line of sight is
$I_{100}/(r_0 n_{\rm H})$.  Both the assumptions of constant $r_0$ and of
constant $n_{\rm H}$ are plausible for a diffuse cloud such as G236+39.
In general, we expect $r_0$ to increase with increasing depth into a
cloud because the ultraviolet radiation that heats the dust responsible
for the 100 $\mu$m continuum emission is itself attenuated by dust absorption.
However, the extinction to the center of G236+39 is relatively small
($A_{\rm V} \sim 0.3$~mag), and the absence of any correlation between
the $\rm 60\mu m/100\mu m$ intensity ratio and the 100 $\mu$m intensity
argues against any significant variation in the dust temperature
(and thus in $r_0$) within the cloud (Reach et al.\ 1994).  Similarly,
considerations of hydrostatic equilibrium imply that the cloud
is only slightly self-gravitating so that the pressure is not expected
to vary significantly within the cloud.  
Since the total pressure in diffuse clouds is dominated by microturbulent 
rather than thermal motions (van Dishoeck \& Black 1989;
Crawford \& Barlow 1996), and since there is no evidence
for significant gradients in the microturbulent line widths, the absence
of significant variations in the pressure implies that the total 
mass density (or equivalently, the density of H nuclei, $n_H$ is roughly
constant).

Our assumption that the 100 $\mu$m intensity
at each position is proportional to the spatial extent of the cloud along
the line of sight does not determine the three-dimensional
shape of the cloud uniquely.  Because the 100 $\mu$m intensity measures
only the extent of the cloud along the line of sight but not its location,
a given 100 $\mu$m map is consistent with a whole set of cloud shapes
that are related by shears perpendicular to the plane of the sky.
Thus we specify a unique cloud shape be making the additional (and
admittedly arbitrary) assumption that the cloud possesses reflection
symmetry about the plane of the sky.

While all our three-dimensional models for G236+39 are based upon the
assumption that the cloud thickness along the line of sight is proportional
to the measured 100 $\mu$m intensity, we have varied the {\it constant
of proportionality} in that relationship so as to obtain the best
fit to the data.  Each model may be characterized conveniently by an
elongation factor $F_E$ which describes the degree of elongation
along the line of sight:
$$F_E \equiv {\sqrt{3\pi} \over 4} {t_{\rm max} \over A^{1/2}}, \eqno(3)$$
where $t_{\rm max} = I_{100}[{\rm max}]/(r_0 n_{\rm H})$ is the maximum
assumed thickness of the cloud, $I_{100}[{\rm max}]$ is the maximum
100 $\mu$m intensity, and $A$ is the projected area of the half-maximum
intensity contour, $I_{100} = 0.5\,I_{100}[{\rm max}]$.  The factor
$\sqrt{3\pi}/4$ has been introduced 
so that the elongation factor $F_E$ is defined
as unity for the case of a sphere; elongation factors smaller than
unity then apply to oblate shapes that are flattened along the line of sight
while elongation factors greater than unity apply to prolate shapes
that are elongated along the line of sight.  In this notation, the
one-dimensional geometry traditionally used to model molecular clouds
is described by the limiting case $F_E=0$.

For G236+39, the angular area of the half-maximum 100 $\mu$m intensity
contour is $7.6 \times 10^{-5} \rm \, sr$, corresponding to a physical
area $A=9.1 \times 10^{37} \, (D/\rm 100\,pc)^2 \, cm^2$ for an assumed
distance to the source of $D$.  The elongation factor is then
related to the density $n_{\rm H}$ by the expression
$$ F_E = 0.56\,\biggl({10^{-20} {\rm MJy\,sr^{-1} cm^2} \over r_0} \biggr)
           \,\biggl({{\rm 100\,pc} \over D} \biggr)
	   \,\biggl({{\rm 100\,cm^3} \over n_{\rm H}} \biggr). \eqno(4)$$

\subsection{$\chi^2$ analysis}

Given the observed IRAS 100 $\mu$m map for G236+39, the molecular
fraction can be predicted at every point in the cloud using the
methods described in \S 2 above.  The results depend upon three
parameters: the 100 $\mu$m emission coefficient per H nucleus, $r_0$;
the elongation factor, $F_E$; and the quantity $I_{UV}/(\alpha n_{\rm H})$,
which represents the relative rates of H$_2$ photodissociation and
H$_2$ formation (see \S  2 above).  Once predictions have been
obtained for the H$_2$ abundance everywhere in the cloud, a predicted
HI column density map can be obtained by integrating $n({\rm H})=
n_{\rm H} - 0.5\,n({\rm H}_2) $ along each line of sight.

We use a standard $\chi^2$ analysis to compare the predicted HI map
with the observations of Reach et al.\ (1994) and thereby to
constrain the values of $r_0$, $F_E$, and
$I_{UV}/(\alpha n_{\rm H})$.
Assuming that the HI observations have a {\it fractional} error that is
roughly the same for each line of sight (Reach et al.\ 1994), the
$\chi^2$ parameter is proportional to the sum
$$ S = \sum \biggl[ (N_{\rm obs}/N_{\rm com}) - 1 \biggr]^2 \eqno(5)$$
over all lines of sight, where $N_{\rm obs}$ is the observed
HI column density and $N_{\rm com}$ is the computed (i.e.\ predicted)
value.

To assess whether three-dimensional (i.e. $F_E > 0$)  models
for G236+39 are superior to one-dimensional ($F_E=0$) models,
we have constructed a series of models of increasing elongation
factor, $F_E$.  For each such model, we have varied the other
parameters, $r_0$ and $I_{UV}/(\alpha n_{\rm H})$, so as to
minimize $\chi^2$.  
Figure 3 
shows the smallest value of
$\chi^2$ that can be achieved for fixed elongation factor, as a function
of $F_E$, relative to the minimum value
that can be achieved for any elongation factor, $\chi^2_{\rm min}$.

Figure 3 shows clearly that the global minimum in $\chi^2$ is obtained
for non-zero $F_E$.  The best estimate of $F_E$ is 0.9, and the
standard error on that estimate is 0.1.  Given the statistical properties
of the $\chi^2$ parameter, the results show at the $60\sigma$ level
that a three-dimensional model with $F_E=0.9$ provides a better fit to
the data than a one-dimensional ($F_E=0$) model.

The comparison between the 3-D ($F_E=0.9$) and 1-D ($F_E=0$) models
can be illustrated graphically in at least three other ways.
In Figure 4, we show the histogram of $N_{\rm obs}/N_{\rm com}$ values
for all 4421 lines of sight; the 3-D model clearly yields a
narrower distribution.  Figure 5 shows the values of $N_{\rm obs}/N_{\rm com}$
as a function of $N_{\rm obs}$.  Again, the 3-D model
(upper panel) clearly shows a tighter
fit to the data, and another feature of the results becomes apparent:
the 1-D model (lower panel) systematically underpredicts the HI column density
(i.e.\ overpredicts the molecular fraction) when
$N({\rm HI}) \sim 3 - 5 \times 10^{20} \rm cm^{-2}$
and overpredicts the HI
column density when $N({\rm HI}) \sim 5 - 8 \times 10^{20} \rm cm^{-2}$.
This behavior is also exhibited by the false color figures displayed in
Plate 1.  Here the two left panels show the value of $N_{\rm obs}/N_{\rm com}$
for every line of sight toward G236+39, the upper left panel applying to
the 1-D model and the lower left panel to the 3-D model.  Values of
$N_{\rm obs}/N_{\rm com}$ close to unity appear green in this figure while
values greater than unity appear yellow or red and values smaller
than unity appear blue or violet.  Again, the 3-D model yields
a better fit to the data, and
the 1-D model shows a systematic tendency to underestimate the HI column
density (and overestimate the molecular fraction) near the edges of
the cloud, where the HI column density lies in the range
$3 - 5 \times 10^{20} \rm cm^{-2}$.  This tendency arises because
the 1-D model does not include the photodissociative effects
of radiation propagating in the plane of the sky; near
the edges of the cloud, such radiation is attenuated by an
H$_2$ column density that is considerably less than the
column density along the line of sight, so the 1-D model
significantly underestimates the photodissociation rate.
The right hand panels in Plate 1 compare the observed ratio of the
100 $\mu$m continuum intensity to the HI column density (lower right panel)
with the predicted ratios for the best fit 3-D model (middle right panel)
and the best fit 1-D model (upper right panel).  The ratio of
the 100 $\mu$m continuum intensity to the HI column density
is a measure of the mean molecular fraction along the line of sight;
once again, the 3-D model clearly provides a better fit.

\subsection{Derived cloud parameters}

The parameters derived for the best fit 1-D model (with the
elongation factor $F_E$ fixed at zero) and the best
fit 3-D model (with $F_E$ taking whatever value optimizes the fit to
the data) are tabulated in Table 1.  Even the 3-D models do not
completely determine the cloud parameters, because the $\chi^2$
analysis places constraints upon three variables ($r_0$, $F_E$,
and $I_{UV}/[\alpha n_H]$) whereas the total number of unknown
parameters is five ($r_0$, $I_{UV}$, $n_{\rm H}$, $\alpha$ and $D$).
Adopting ``canonical values" for the ultraviolet radiation
field ($I_{UV}=1$) and for the H$_2$ formation rate coefficient
($\alpha=1$), we obtain estimates of 
$(1.13\pm 0.06)\times 10^{-20} \rm \,MJy\,sr^{-1} cm^2$ for the 100 $\mu$m
emission coefficient per H nucleus, $r_0$; $(48 \pm 7) \rm\, cm^{-3}$ 
for the density of H nuclei, $n_{\rm H}$; 
and $(110 \pm 20)$~pc for the distance to the source, $D$.  (All the
error bounds given here are $1\sigma$ statistical uncertainties.) 
Figure 6 shows the three-dimensional shape of the cloud G236+39 
that is implied by the elongation factor that we derived.  

\section{Discussion}

The $\chi^2$ analysis described in \S 3.2 above
establishes convincingly that the abundance of H$_2$ in G236+39
is explained more successfully by a 3-D model than by a 1-D model.
Furthermore, the 3-D model provides a distance estimate of 
$D \sim 100$~pc which is at least consistent with the known scale
height of cold HI in the Galaxy.

\subsection{Validation of the model}

As a further test of our method, we have applied it to a 
set of artificial 100 $\mu$m and HI 21 cm maps that we
generated for the case of a triaxial spheriod of elongation
factor unity.  After adding artificial noise to those maps,
we attempted to reconstruct the elongation factor for
that spheriod using a $\chi^2$ analysis exactly
analogous to that described in \S 3.2.  The elongation
factor was indeed recovered successfully, the derived 
value being $1.02 \pm 0.06\,\, (1\sigma)$.  Full details
of this analysis and figures analogous to those described
in \S 3 are presented in the Appendix.

\subsection{Limitations of the model}

Despite its apparent success in providing a better fit
to the observational data than a 1-D model, our 3-D model
is clearly idealized in that we have made several simplifying
assumptions.  The assumptions of constant $n_{\rm H}$ and $r_0$
have already been discussed in \S 3.1 above.  Our assumption
that the cloud G236+39 shows reflection symmetry about the
plane of the sky has also been discussed; because we
assume that all the low-level 100 $\mu$m emission near the edges of the
cloud arises from material that is located close to the symmetry
plane, this assumption probably leads to
an overestimate of the degree of shielding at the cloud center.
A further simplifying assumption is our adoption of an isotropic
incident radiation field.

We have also assumed that the H$_2$ chemistry is in equilibrium, with the
rate of photodissociation equal to the rate of formation.
The timescale for establishing such a chemical equilibrium is
$\tau_{\rm chem} = 1/(R n_{\rm H}) \sim 20 \rm \, Myr$.  (Note that
the estimate given by Reach et al.\ 1994 is too large by an order of magnitude.)
For comparison, we may define a crossing timescale as the ratio
of the cloud size to the HI linewidth, $\tau_{\rm cross} = I_{100}[{\rm max}]/
(n_{\rm H} r_0 \Delta v) \sim 3 \rm \,Myr$, where $\Delta v \sim 3 \rm \, km\,s^{-1}$
is the typical HI linewidth.  The fact that $\tau_{\rm chem}$
exceeds $\tau_{\rm cross}$
does not necessarily invalidate our assumption of chemical equilibrium.
Even assuming that turbulent motions contribute
significantly to the line width, we would expect that turbulent mixing
will tend to decrease the molecular fraction in the cloud interior only if
$\tau_{\rm chem}$ also exceeds the turbulent diffusion timescale,
$\tau_{\rm diff}$, which we expect to be larger than $\tau_{\rm cross}$
by the ratio of the cloud size to the effective mixing length.  Turning
the argument around, we suggest that the success of our equilibrium 
models in explaining the observations implies that $\tau_{\rm diff}$ is indeed
{\it long} relative to $\tau_{\rm chem}$ and that the ratio of the cloud 
size to the effective mixing length is greater than $\sim 10$.

\subsection{Future studies of H$_2$ in diffuse molecular clouds}

We anticipate several future extensions to the study presented in this
paper.  Thus far, we have only applied our 3-D modeling methods
to the single source G236+39; one obvious extension will be their
application to other isolated diffuse clouds. In addition, we
expect that observations at other wavelengths will provide
useful additional probes of the physical and chemical conditions
in diffuse clouds such as G236+39.   In particular, CII fine structure 
emissions  in the $\rm ^3P_{3/2} - ^3P_{1/2}$ line near 158 $\mu$m would
provide a valuable independent probe of the density
and a check on our assumption that $n_{\rm H}$ is roughly
constant.   Except at the cloud
center, from which CO emissions have been detected (Reach et al.\ 1994),
we expect C$^+$ to be the dominant reservoir of gas-phase carbon
nuclei in G236+39; thus C$^+$ will have a roughly
constant abundance relative to H nuclei.
Because the density in diffuse clouds is very much smaller
than the critical density for the CII $\rm ^3P_{3/2} - ^3P_{1/2}$
transition, the frequency-integrated CII line intensity 
will therefore be proportional to $N_{\rm H} n_{\rm H}$,
and the CII line equivalent width will be proportional to $n_{\rm H}$.
The {\it Infrared Space Observatory} (ISO), the {\it Stratospheric
Observatory for Infrared Astronomy} (SOFIA), as well as planned
balloon-borne telescopes will provide ideal platforms for carrying
out CII observations of diffuse clouds.  We note 
that only a relatively small telescope (with a 20~cm diameter
primary mirror) is needed to provide a beam size at
158 $\mu$m that matches the beam size of the Arecibo radiotelescope
for H 21 cm observations. 

The far ultraviolet is another spectral region that is
important for the study of molecular
hydrogen in diffuse clouds.  In particular,
the launch of the {\it Far Ultraviolet Spectroscopic
Explorer} will permit the direct measurement of H$_2$ column densities
in absorption
along specific lines of sight toward ultraviolet continuum sources.
Although the spatial sampling provided by such absorption line
studies will be limited by the availability of bright
background sources, FUSE observations will provide a very
valuable complement to the indirect methods for studying H$_2$
that have been described in this paper.

We are very grateful to W.~Reach for making available the H 21cm map of 
G236+39, and to E.~van~Dishoeck for allowing us to use her
computer code as a check on our results in the limit $F_E=0$.
We acknowledge with gratitude the support of NASA grant NAGW-3147 from the
Long Term Space Astrophysics Research Program.
Most of the simulations presented in this work were performed on the Cray Y-MP
operated by the Netherlands Council for Supercomputer Facilities in Amsterdam.
\newpage
\centerline{APPENDIX}

As a further test of our method, we have applied it to a 
set of artificial 100 $\mu$m and HI 21 cm maps that we
generated for the case of a triaxial spheriod.

We considered a spheriod with an axial ratio 1 : $\sqrt{2}$ : 2, with
the longest and shortest
axes in the plane of the sky, and
the intermediate-length axis oriented along the line of sight.
The elongation factor is therefore $F_E=1$.  We took a peak 
$I_{100}$ $\mu$m
intensity of 10 MJy sr$^{-1}$, an emission coefficient
per H nucleus of $r_0=1.0\times 10^{-20}$ MJy sr$^{-1}$ cm$^2$, 
a gas density $n_{\rm H}=50$ cm$^{-3}$, and an isotropic incident
UV field $I_{\rm UV}=1$.  We generated artificial 100 $\mu$m 
and HI 21 cm maps for this object, adding Gaussian noise to each
on the scale of the HI beam (3 arcminutes) with a FWHM of 7\%. 
The algorithm described in the main text was then used to obtain 
the best fit 3-D model for these artificial maps, in an attempt
to recover the input parameters.  The results are presented in 
Figures A1, A2, A3, A4, and Plate 2 (which
correspond directly to Figures 1, 3, 4, 5, and Plate 1 discussed
previously for the case of G236+39.)
 
Figure A1 presents the artificial data and a curve corresponding
to the best fit
1-D model. Figure A2 shows minimum
$\chi^2$ value attained at fixed elongation
factor, as a function of $F_E$.
As in the case of the G236+39 data, the overall best fit
is found for a model with non-zero $F_E$.  Indeed, our $\chi^2$
analysis successfully recovers the input values for the 
elongation factor [$F_E = 1.02 \pm 0.06 \,(1 \sigma)$];
as well as for $r_0$ [$=1.01\pm 0.03\,(1 \sigma)$ $10^{-20}$ 
MJy sr$^{-1}$ cm$^2$] and for $I_{UV}$ [$=1.02 \pm 0.06\,(1 \sigma)$].

Figure A3 compares the histogram of $N_{\rm obs}/N_{\rm com}$ values
for the 1-D and 3-D models (c.f. \S 3.2), 
and Figure A4 shows the values of $N_{\rm obs}/N_{\rm com}$
as a function of $N_{\rm obs}$.  
As in the case of the G236+39 data, the 3-D model
(upper panel) clearly shows a tighter
fit to the data, and the
the 1-D model (lower panel) systematically underpredicts the HI column density
(i.e.\ overpredicts the molecular fraction) for intermediate
HI column densities.  Plate 2 shows false color images that
are analogous to those in
Plate 1.  Here the two left panels show the value of 
$N_{\rm obs}/N_{\rm com}$
for every line of sight in the artificial map, the upper left panel applying to
the 1-D model ($F_E=0$), and the lower left panel to the best fit 3-D model
($F_E=1$). 
As before (c.f. \S3.2) , the 1-D model shows a systematic tendency to 
underestimate the HI column
density (and overestimate the molecular fraction) near the edges of
the cloud; this effect is reversed when the adopted value of
$F_E$ is too large.  The right hand panels in Plate 2 compare the observed 
ratio of the
100 $\mu$m continuum intensity to the HI column density (lower right panel)
with the predicted ratios for the best fit 3-D model (middle right panel)
and the best fit 1-D model (upper right panel); once again, the
3-D model clearly provides a better fit.

\newpage

\begin{table}[h,b,t]
        \begin{center}
        \caption{Derived cloud parameters for G236+39$^a$}
        \vspace{0.3cm}
        \begin{tabular}{lll} \hline\hline
        & 3-D model & 1-D model \\ \hline
$r_0$   & $(1.13 \pm 0.06) \times 10^{-20} \rm \,MJy\,sr^{-1}\,cm^{2}$ 
        & $(1.01 \pm 0.10) \times 10^{-20} \rm \,MJy\,sr^{-1}\,cm^{2}$ \\
$I_{UV}/(n_{\rm H} \alpha)$
        & $(0.021 \pm 0.003) \rm \, cm^3$ & $(0.018 \pm 0.005) \rm \, cm^3$ \\
$n_{\rm H} \alpha/I_{UV}$ 
        & $(48 \pm 7) \rm \, cm^{-3}$ &  $(57 \pm 18) \rm \, cm^{-3}$ \\
$F_E$   & $0.94 \pm 0.1$ & $0^b$ \\
$I_{UV} D /\alpha $
        & $(110 \pm 20)$~pc & \\
$n_{\rm H} (D/ 100\,\rm pc)$ 
        &  $(53 \pm 8) \rm \, cm^{-3}$ \\
        \hline\hline
        \end{tabular}
        \end{center}

$^a$ All error bounds are 1$\sigma$ statistical errors

$^b$ $F_E=0$ by assumption for the 1-D model

\end{table}

\newpage

\centerline{FIGURE CAPTIONS}

\figcaption{Correlation between 100 $\mu$m continuum intensity and
HI column density for 4421 lines of sight toward the IRAS cloud G236+39.
The HI data are from Reach et al.~(1994, their Figure 12). The solid line corresponds to the best fit one-dimensional model described in the text.}

\figcaption{Comparison between the H$_2$ and H abundances obtained 
using the 1-D computer code of Black \& van Dishoeck
(1987, dotted) and those obtained using our 3-D computer code
(solid) for the case of a plane parallel slab with
$T=50$ K, $I_{\rm UV}=3$, $n_{\rm H}=200$ cm$^{-3}$ and $A_{\rm V}=1$ mag.}

\figcaption{$\chi^2$ analysis for HI observations of the
IRAS cloud G236+39.  The curve shows the smallest value of
$\chi^2$ that can be achieved for fixed elongation factor, as a function
of $F_E$, relative to the minimum value
that can be achieved for any elongation factor, $\chi^2_{\rm min}$.}

\figcaption{Histogram of values for the ratio of observed to
predicted HI column density, $N_{\rm obs}/N_{\rm com}$,
for all 4421 lines of sight in G236+39.  The relative frequency on
the vertical axis is normalized such that the total area under
the curve is unity.}

\figcaption{Values for the ratio of observed to
predicted HI column density, $N_{\rm obs}/N_{\rm com}$,
for all 4421 lines of sight in G236+39, as a function of
observed HI column, $N_{\rm obs}$.}

\figcaption{Shape of IRAS G236+39 implied by the best-fit 3-D model.
The vertical scale has been chosen so that the extent of the cloud
along the line of sight is represented accurately relative to
the scale in the plane of the sky, given the elongation
factor derived from our $\chi^2$ analysis (\S3.2).
Note that Figure 6 shows only
the upper half of the cloud; the lower half is assumed to be an
exact reflection of the upper half in the plane of the sky.}

{\parindent 0pt

Fig. A1.--- Correlation between 100 $\mu$m continuum intensity and
HI column density for the artificial data set discussed in the Appendix.
The solid line corresponds to the best fit one-dimensional model.

Fig. A2.--- $\chi^2$ analysis for HI column densities in
the artificial data set discussed in the Appendix.
The curve shows the smallest value of
$\chi^2$ that can be achieved for fixed elongation factor, as a function
of $F_E$, relative to the minimum value
that can be achieved for any elongation factor, $\chi^2_{\rm min}$

Fig. A3.--- Histogram of values for the ratio of ``observed" to
predicted HI column density, $N_{\rm obs}/N_{\rm com}$,
for the artificial data set discussed in the Appendix.  The relative 
frequency on the vertical axis is normalized such that the total area 
under the curve is unity.

Fig. A4.--- Values for the ratio of ``observed" to
predicted HI column density, $N_{\rm obs}/N_{\rm com}$,
for the artificial data set discussed in the Appendix, as a function of
``observed" HI column, $N_{\rm obs}$.

Plate 1.--- {\it Left panels}: ratio of
observed to predicted HI column density, $N_{\rm obs}/N_{\rm com}$,
for every line of sight toward G236+39, the upper left panel applying to
the best-fit 1-D model and the lower left panel to the best-fit
3-D model.  The lower left panel also shows contours of $I_{100}$ ranging
from 3 to 11 MJy sr$^{-1}$ in steps of 2 MJy sr$^{-1}$.
{\it Right panels}: comparison of observed ratio of the
100 $\mu$m continuum intensity to the HI column density (lower right panel)
with the predicted ratios for the best fit 3-D model (middle right panel)
and the best fit 1-D model (upper right panel).
The map centers all have coordinates $\alpha =09^h 43^m$ and
$\delta =01^{\circ} 20'$ (1950.0).

Plate 2.--- {\it Left panels}: ratio of
``observed" to predicted HI column density, $N_{\rm obs}/N_{\rm com}$,
for the artificial data set, the upper left panel applying to
the 1-D model and the lower left panel to the best fit 3-D model. {\it Right panels}: comparison of ``observed" ratio of the
100 $\mu$m continuum intensity to the HI column density (lower right panel)
with the predicted ratios for the best fit 3-D model (middle right panel)
and the best fit 1-D model (upper right panel).

}

\setcounter{figure}{0}

\newpage
\begin{figure}
\centerline{\psfig{figure=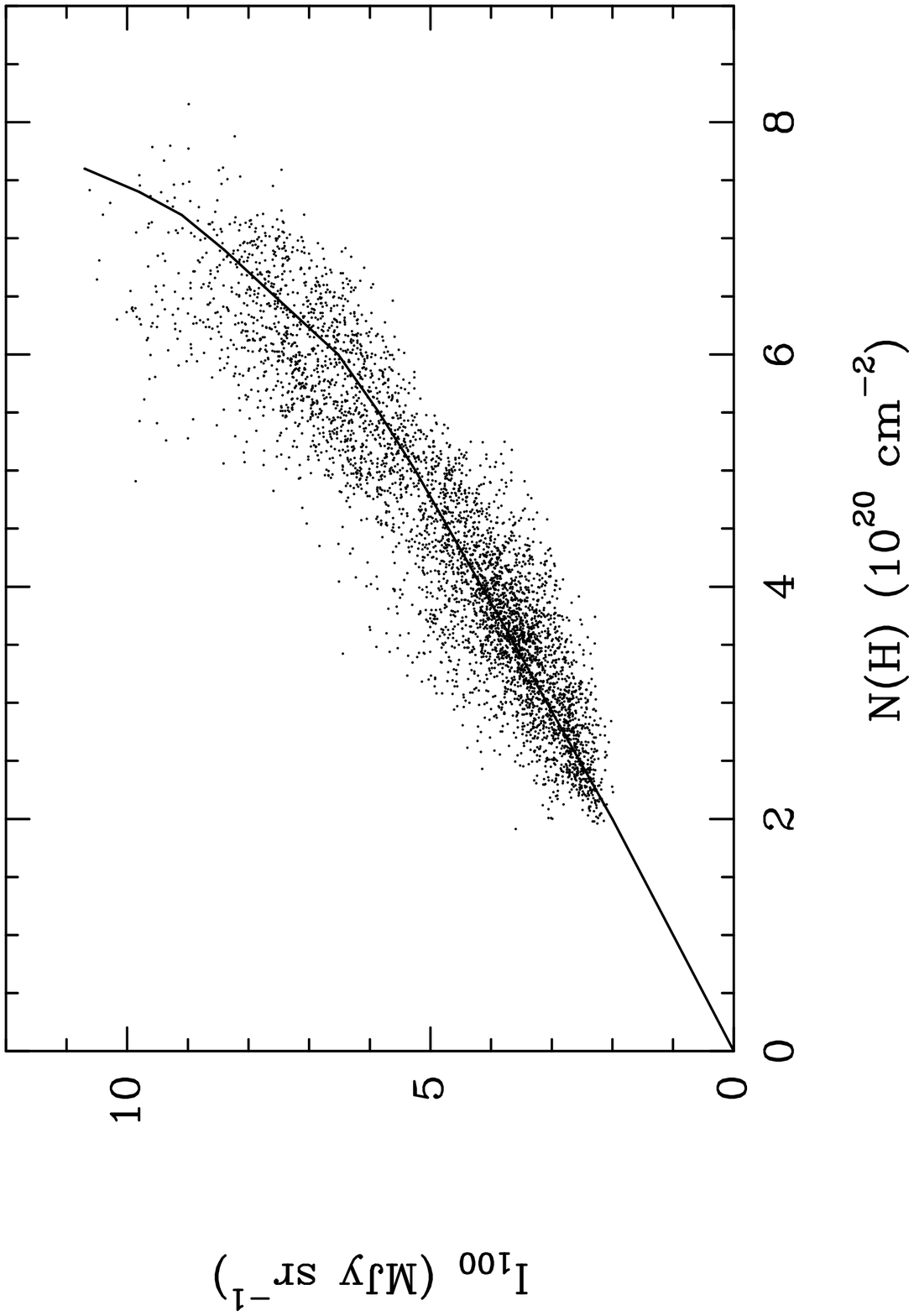,width=6in}}
\caption{}
\end{figure}

\newpage
\begin{figure}
\centerline{\psfig{figure=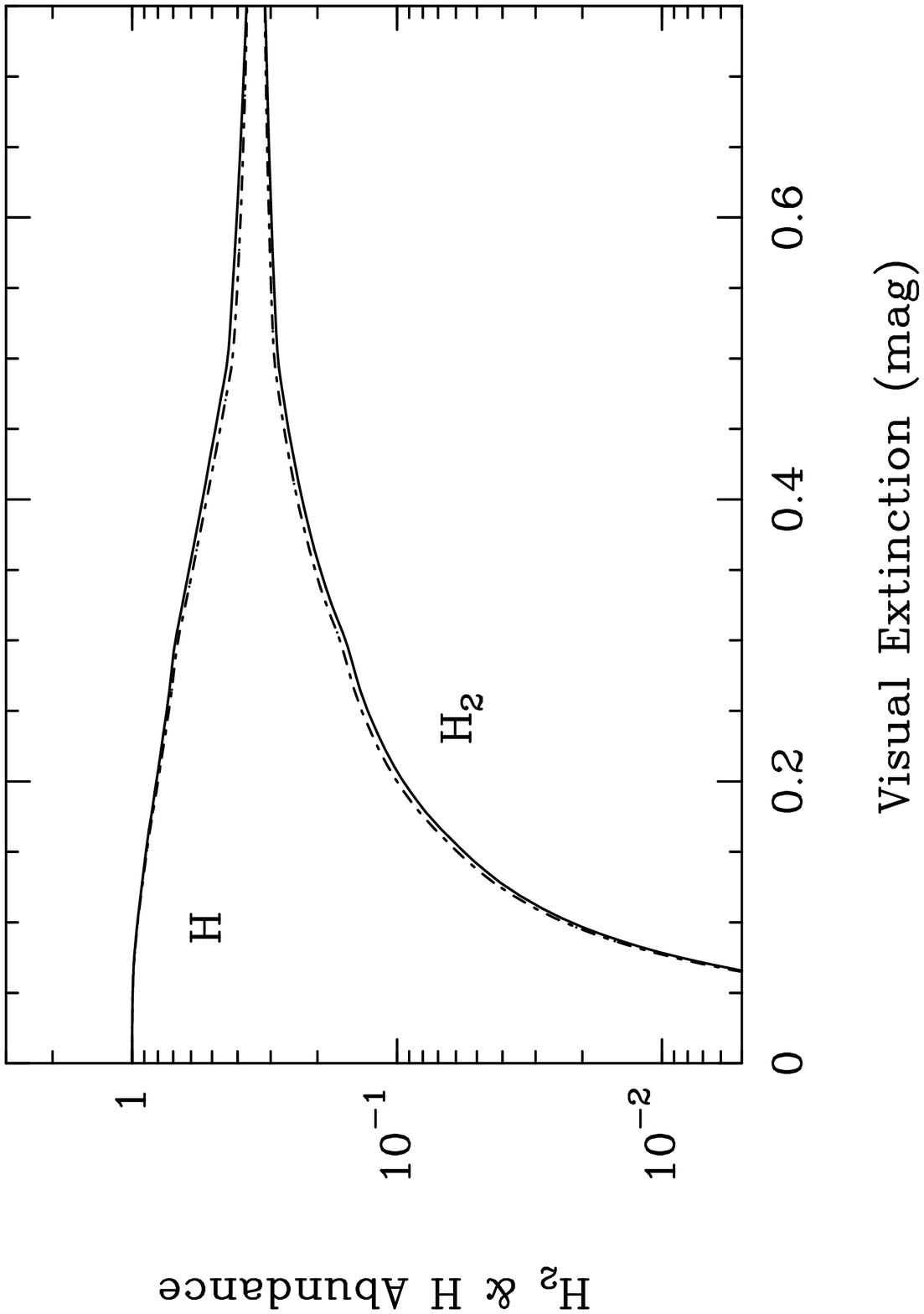,width=6in}}
\caption{}
\end{figure}

\newpage
\begin{figure}
\centerline{\psfig{figure=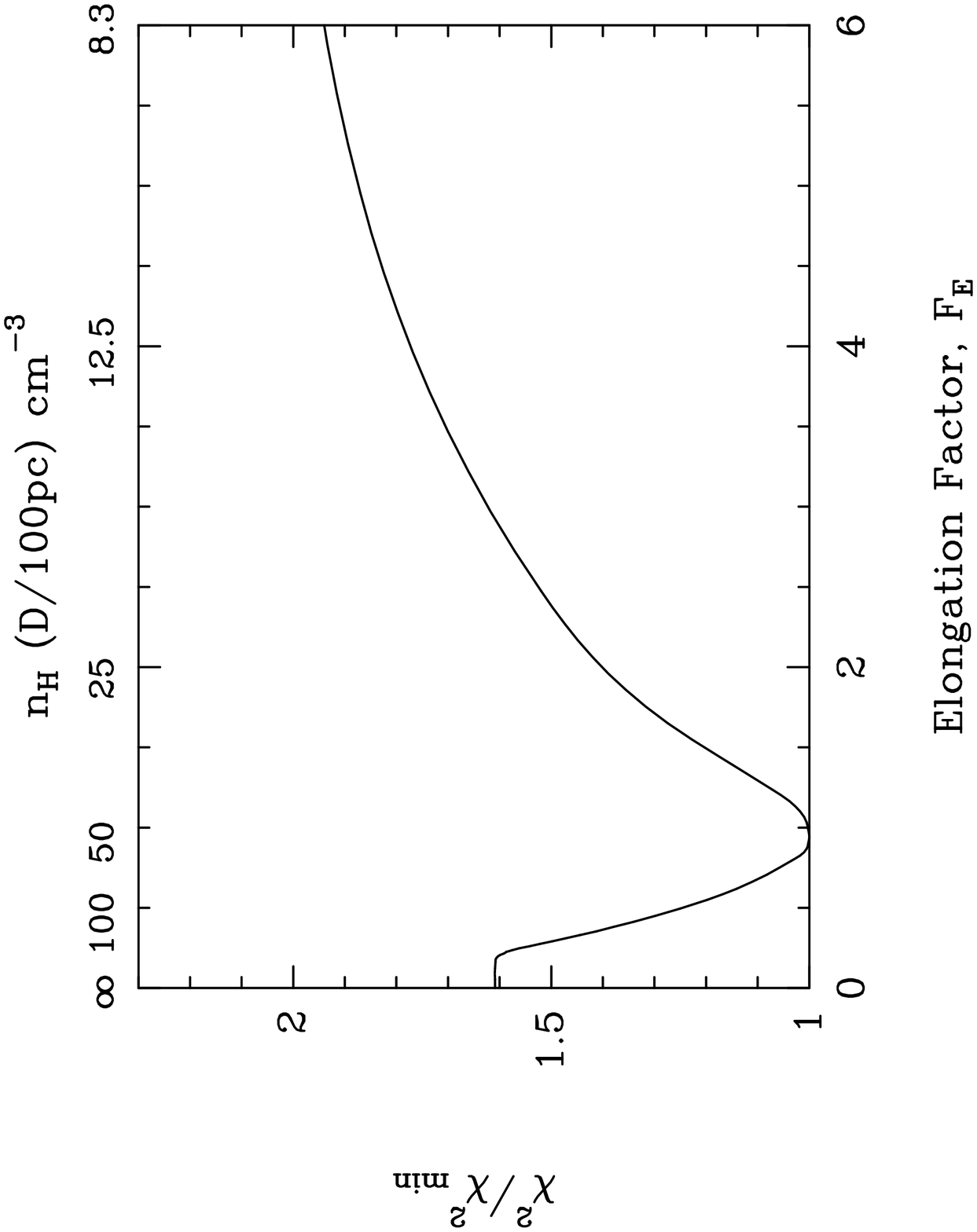,width=6in}}
\caption{}
\end{figure}

\newpage
\begin{figure}
\centerline{\psfig{figure=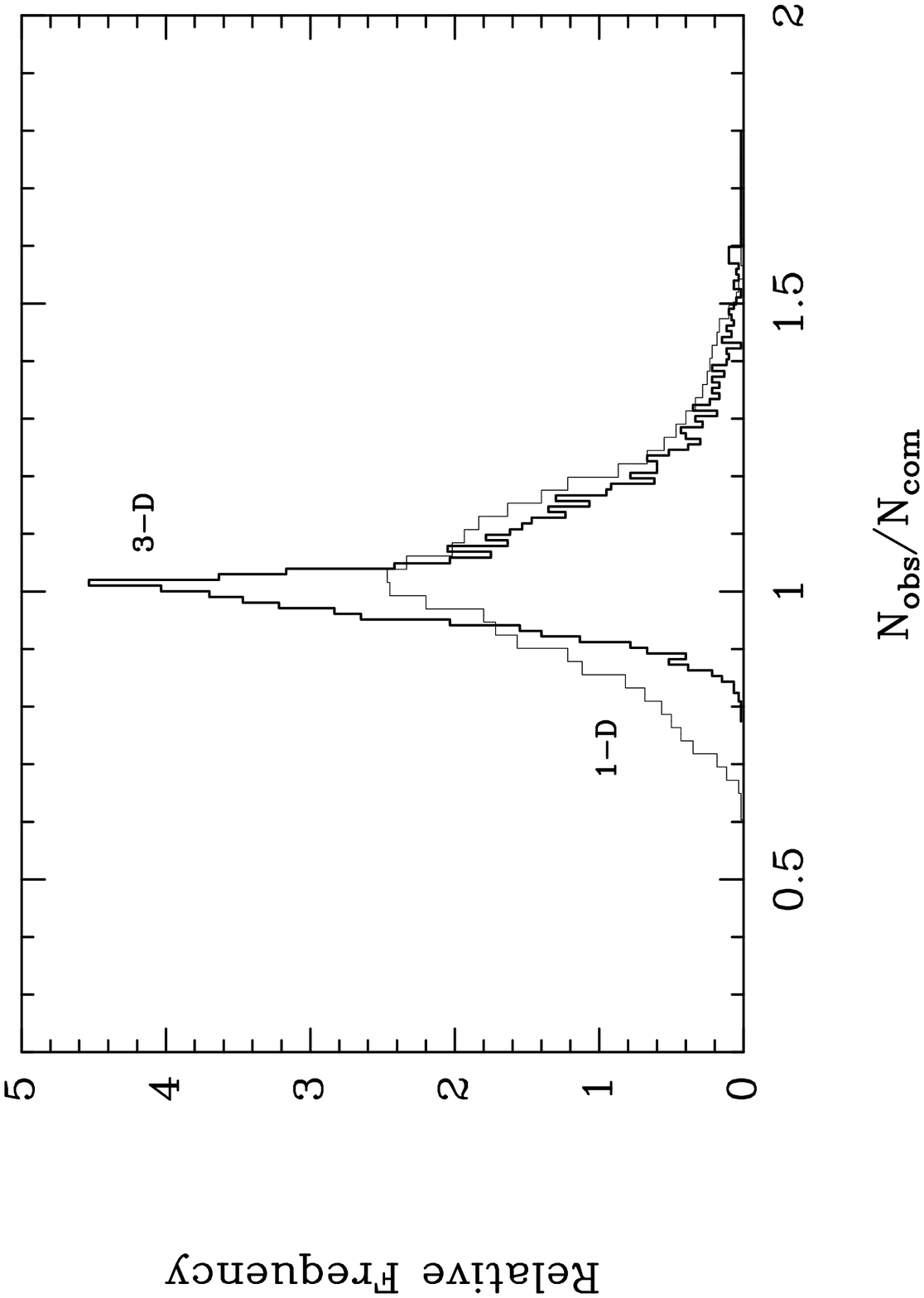,width=6in}}
\caption{}
\end{figure}

\newpage
\begin{figure}
\centerline{\psfig{figure=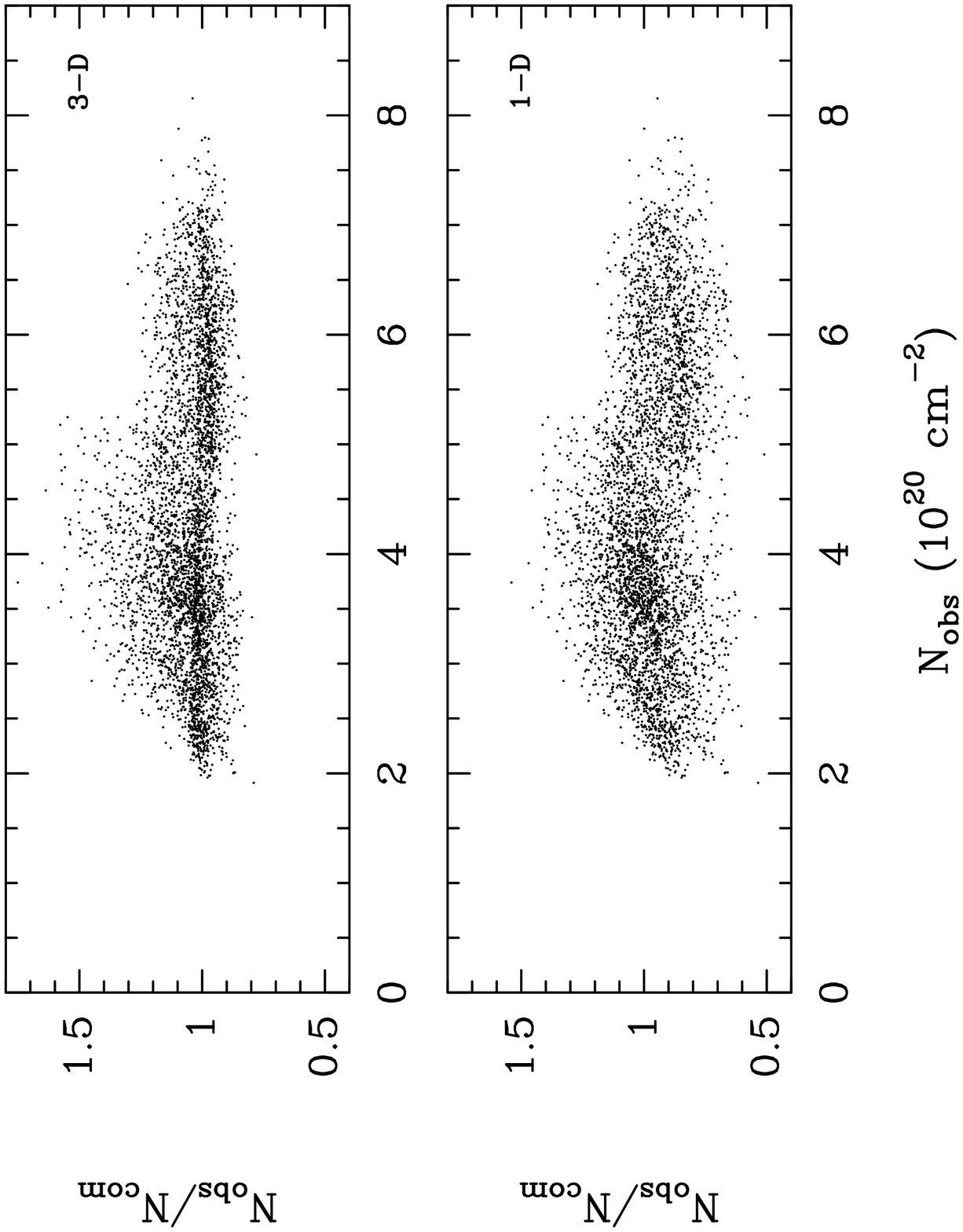,width=6in}}
\caption{}
\end{figure}

\newpage
\begin{figure}
\centerline{\psfig{figure=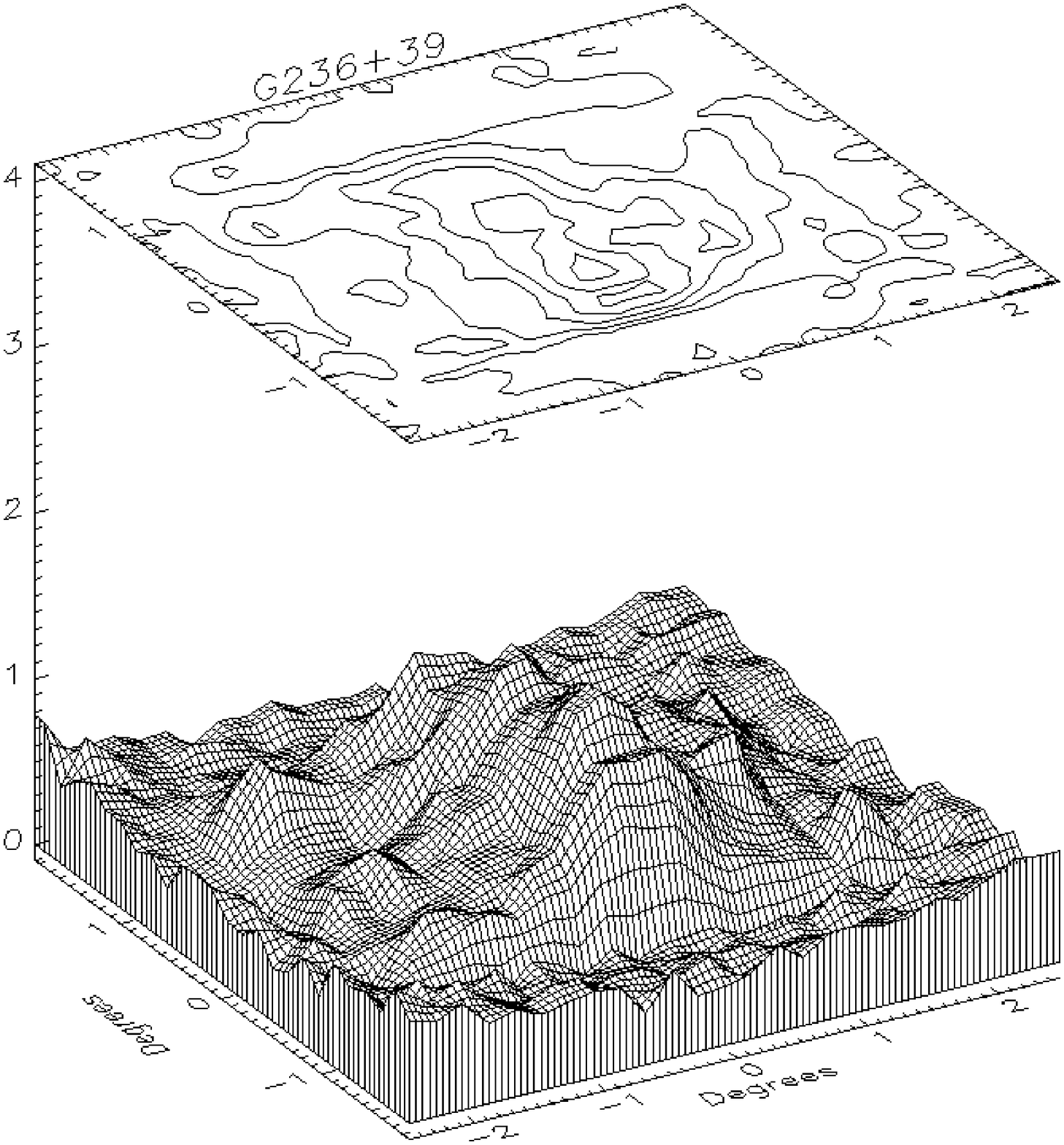,width=6in}}
\caption{}
\end{figure}

\setcounter{figure}{0}

\newpage
\begin{figure}
\centerline{\psfig{figure=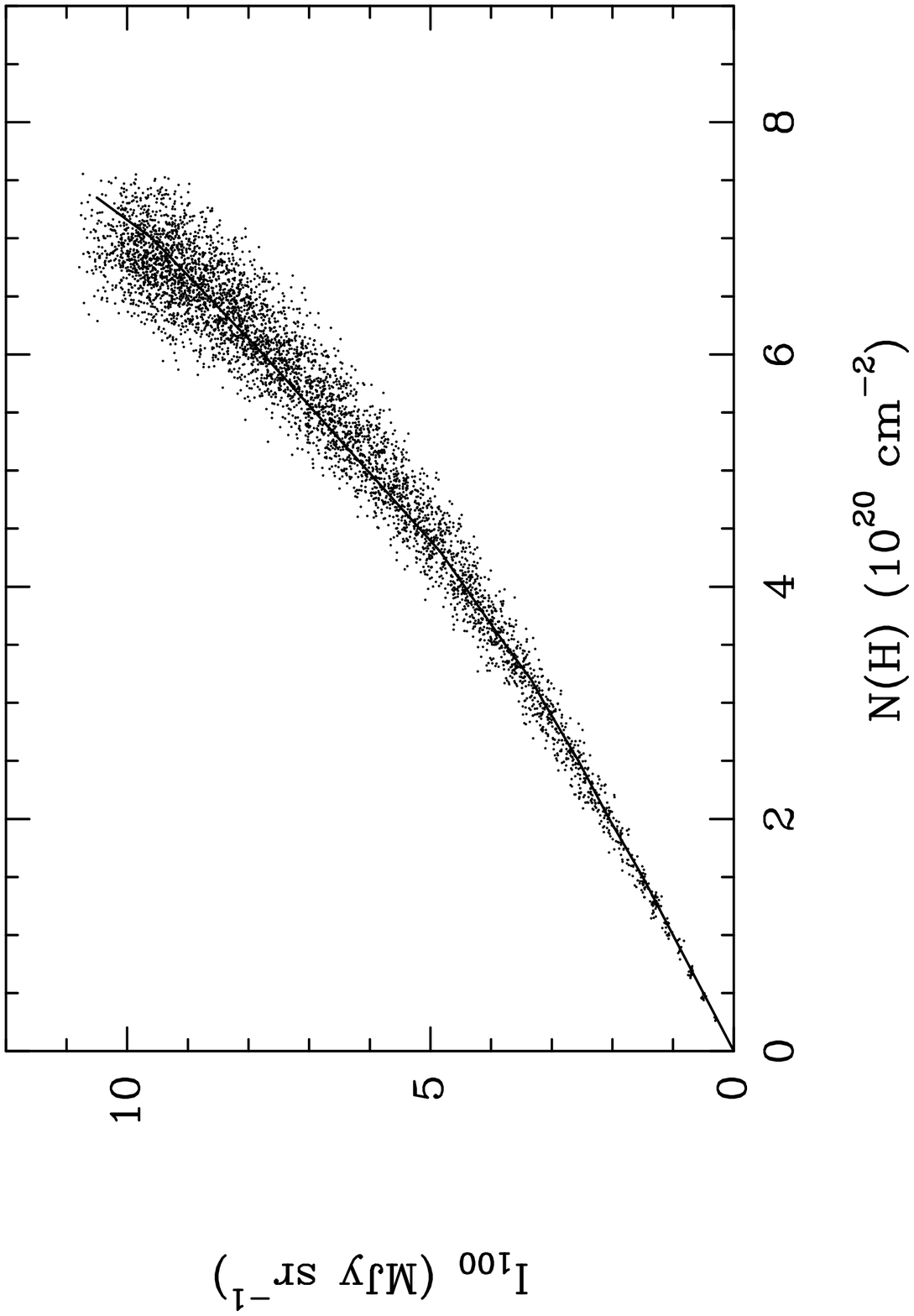,width=6in}}
\caption{Appendix}
\end{figure}

\newpage
\begin{figure}
\centerline{\psfig{figure=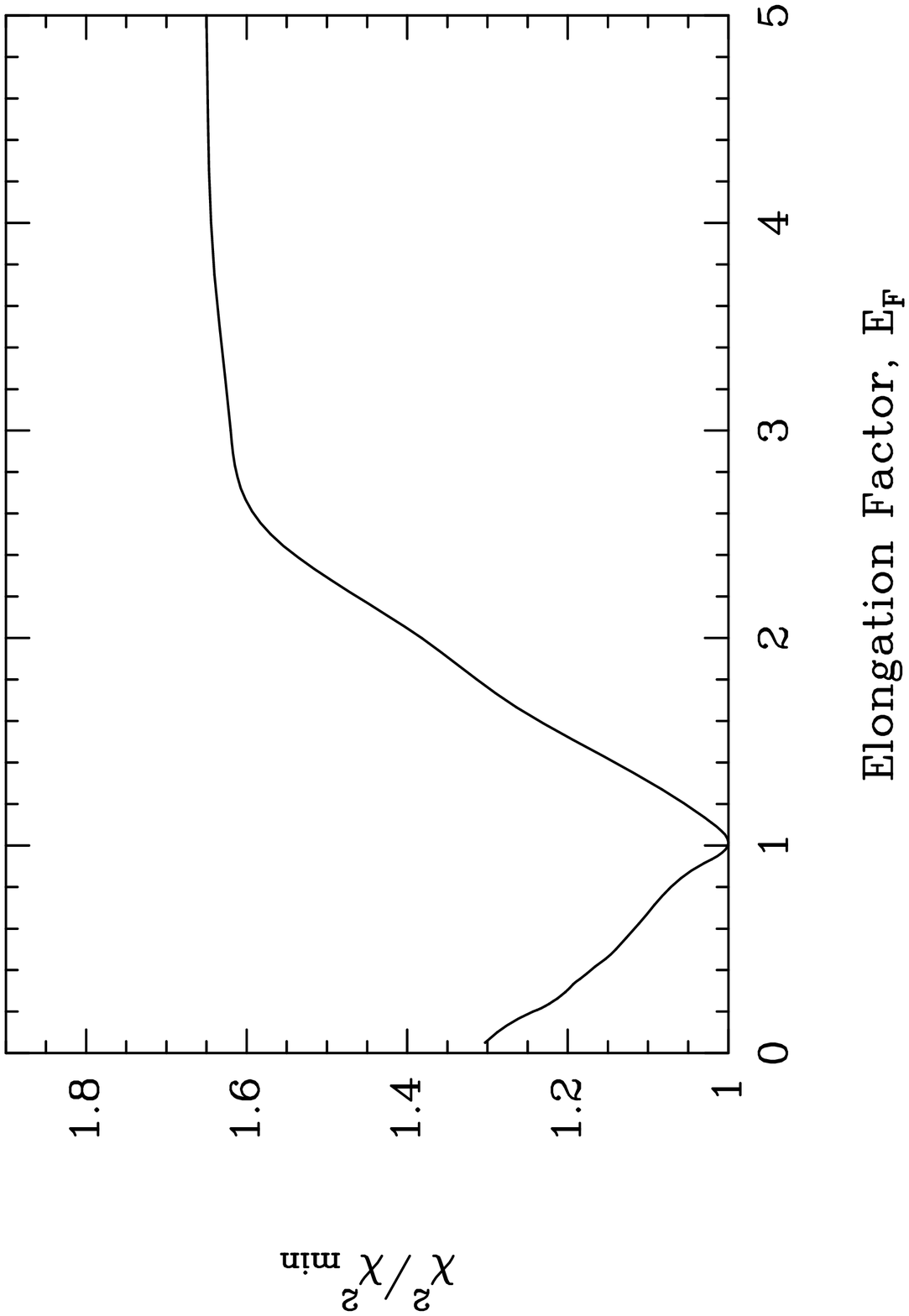,width=6in}}
\caption{Appendix}
\end{figure}

\newpage
\begin{figure}
\centerline{\psfig{figure=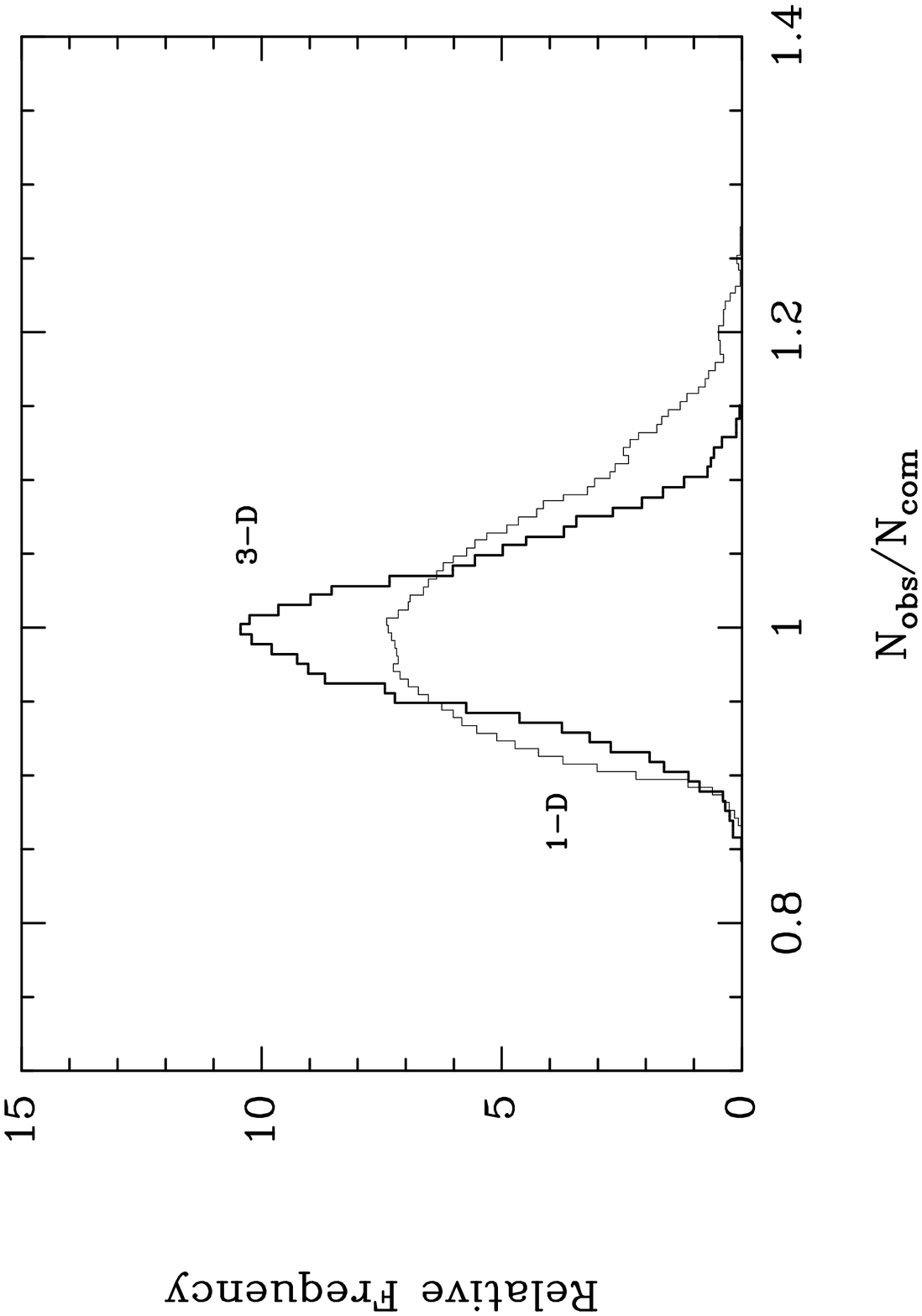,width=6in}}
\caption{Appendix}
\end{figure}

\newpage
\begin{figure}
\centerline{\psfig{figure=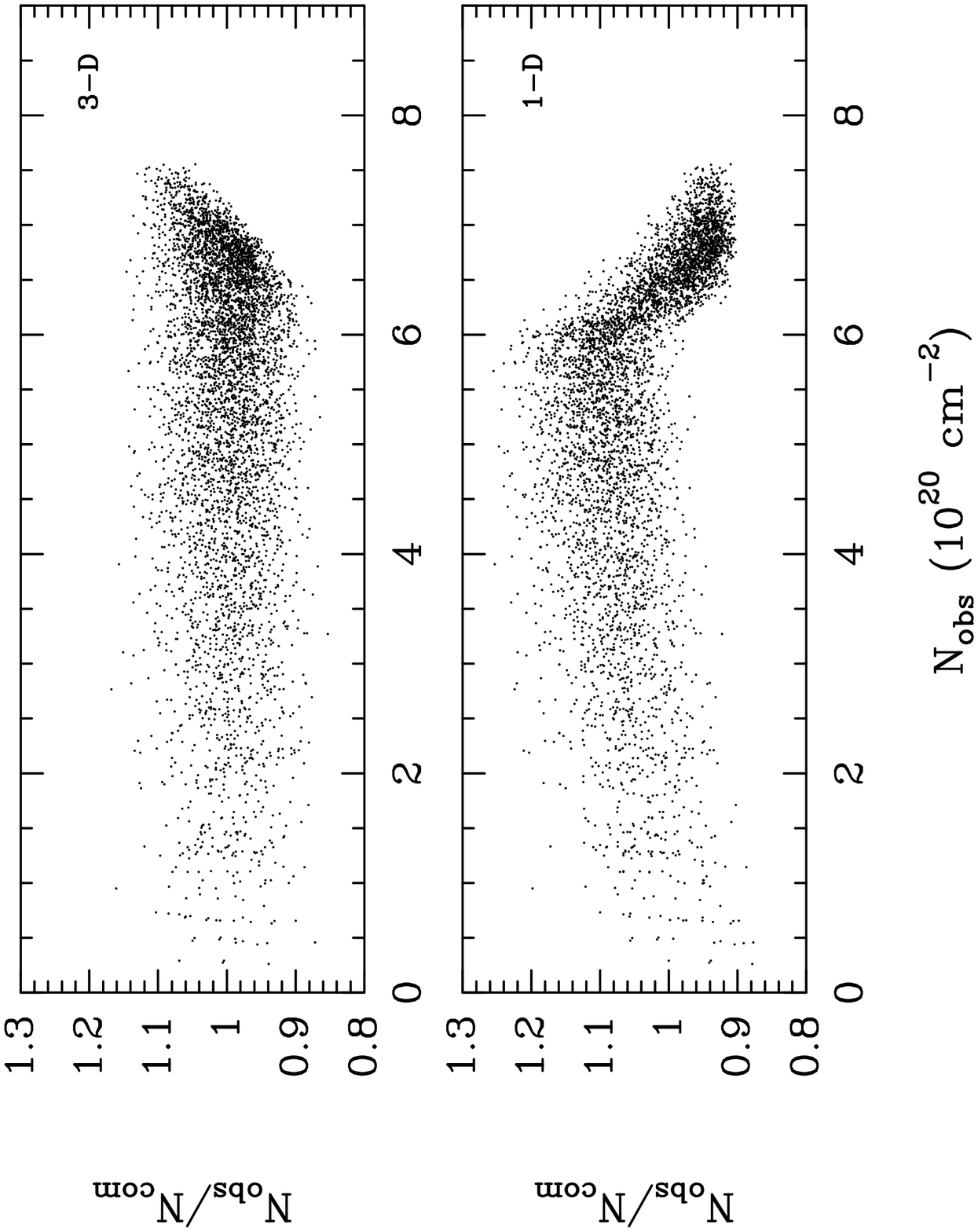,width=6in}}
\caption{Appendix}
\end{figure}

\newpage
\begin{figure}
\centerline{Color Plate 1 may be viewed on the World Wide Web at URL:}
\centerline{http://www.pha.jhu.edu/\~\thinspace neufeld/spaans+neufeld/plate1.html}
\end{figure}
\begin{figure}
\centerline{Color Plate 2 may be viewed on the World Wide Web at URL:}
\centerline{http://www.pha.jhu.edu/\~\thinspace neufeld/spaans+neufeld/plate2.html}
\bigskip\bigskip
\centerline{The Web browser's window should be set as large as possible to
view these large color images.}
\end{figure}

\end{document}